\documentclass[12pt,epsf]{article}

\usepackage{times}
\usepackage{graphicx}
\usepackage{amsfonts}
\usepackage{amsmath}
\usepackage{amssymb}
\usepackage{amstext}
\usepackage{amsthm}
\usepackage{youngtab}

\usepackage{epsfig}
\usepackage{color}
\usepackage{leftidx}
\usepackage{tikz}
\psfigdriver{dvips}
\usepackage{latexsym}
\usepackage[matrix,curve]{xypic}
\usepackage{rotating}
\rotdriver{dvips}

\begin{document}

\baselineskip 6mm
\renewcommand{\thefootnote}{\fnsymbol{footnote}}


\newcommand{\nc}{\newcommand}
\newcommand{\rnc}{\renewcommand}


\rnc{\baselinestretch}{1.24}    
\setlength{\jot}{6pt}       
\rnc{\arraystretch}{1.24}   

\makeatletter
\rnc{\theequation}{\thesection.\arabic{equation}}
\@addtoreset{equation}{section}
\makeatother



\nc{\be}{\begin{equation}}

\nc{\ee}{\end{equation}}

\nc{\bea}{\begin{eqnarray}}

\nc{\eea}{\end{eqnarray}}

\nc{\xx}{\nonumber\\}

\nc{\ct}{\cite}

\nc{\la}{\label}

\nc{\eq}[1]{(\ref{#1})}

\nc{\newcaption}[1]{\centerline{\parbox{6in}{\caption{#1}}}}

\nc{\fig}[3]{
\begin{figure}
\centerline{\epsfxsize=#1\epsfbox{#2.eps}}
\newcaption{#3. \label{#2}}
\end{figure}
}


\def\CA{{\cal A}}
\def\CC{{\cal C}}
\def\CD{{\cal D}}
\def\CE{{\cal E}}
\def\CF{{\cal F}}
\def\CG{{\cal G}}
\def\CH{{\cal H}}
\def\CK{{\cal K}}
\def\CL{{\cal L}}
\def\CM{{\cal M}}
\def\CN{{\cal N}}
\def\CO{{\cal O}}
\def\CP{{\cal P}}
\def\CR{{\cal R}}
\def\CS{{\cal S}}
\def\CU{{\cal U}}
\def\CV{{\cal V}}
\def\CW{{\cal W}}
\def\CY{{\cal Y}}
\def\CZ{{\cal Z}}


\def\IB{{\hbox{{\rm I}\kern-.2em\hbox{\rm B}}}}
\def\IC{\,\,{\hbox{{\rm I}\kern-.50em\hbox{\bf C}}}}
\def\ID{{\hbox{{\rm I}\kern-.2em\hbox{\rm D}}}}
\def\IF{{\hbox{{\rm I}\kern-.2em\hbox{\rm F}}}}
\def\IH{{\hbox{{\rm I}\kern-.2em\hbox{\rm H}}}}
\def\IN{{\hbox{{\rm I}\kern-.2em\hbox{\rm N}}}}
\def\IP{{\hbox{{\rm I}\kern-.2em\hbox{\rm P}}}}
\def\IR{{\hbox{{\rm I}\kern-.2em\hbox{\rm R}}}}
\def\IZ{{\hbox{{\rm Z}\kern-.4em\hbox{\rm Z}}}}


\def\a{\alpha}
\def\b{\beta}
\def\d{\delta}
\def\ep{\epsilon}
\def\ga{\gamma}
\def\k{\kappa}
\def\l{\lambda}
\def\s{\sigma}
\def\t{\theta}
\def\w{\omega}
\def\G{\Gamma}


\def\half{\frac{1}{2}}
\def\dint#1#2{\int\limits_{#1}^{#2}}
\def\goto{\rightarrow}
\def\para{\parallel}
\def\brac#1{\langle #1 \rangle}
\def\curl{\nabla\times}
\def\div{\nabla\cdot}
\def\p{\partial}


\def\Tr{{\rm Tr}\,}
\def\det{{\rm det}}


\def\vare{\varepsilon}
\def\zbar{\bar{z}}
\def\wbar{\bar{w}}


\def\ad{\dot{a}}
\def\bd{\dot{b}}
\def\cd{\dot{c}}
\def\dd{\dot{d}}
\def\so{SO(4)}
\def\bfr{{\bf R}}
\def\bfc{{\bf C}}
\def\bfz{{\bf Z}}

\begin{titlepage}


\hfill\parbox{3.7cm} {{\tt arXiv:2409.14808}}

\vspace{15mm}

\begin{center}
{\Large \bf Exact Results On the Number of Gravitons \\ Radiated During Binary Inspiral}

\vspace{10mm}

Youngjoo Chung ${}^{a}$\footnote{ychung@gist.ac.kr} and Hyun Seok Yang ${}^b$\footnote{hsyang@gist.ac.kr}
\\[10mm]

${}^a$ {\sl School of Electrical Engineering and Computer Science, \\ Gwangju Institute of Science and Technology, Gwangju 61005, Korea}

${}^b$ {\sl Department of Physics and Photon Science, \\ Gwangju Institute of Science and Technology, Gwangju 61005, Korea}

\end{center}

\thispagestyle{empty}

\vskip1cm


\centerline{\bf ABSTRACT}
\vskip 4mm
\noindent

We derive an exact formula $F(e)$ which provides a concrete estimate for the total number and angular momentum of gravitons emitted during the nonrelativistic inspiral of two black holes.
We show that the function $F(e)$ is a slowly growing monotonic function of the eccentricity $0 \le e \le 1$
and $F(1) = 1.0128 \cdots $.
We confirm and extend the results obtained by Page for the function $F(e)$. We also get an exact result for the ratio
$\nu (e_i) = \frac{2\hbar N(L_i, e_i)}{L_i}$ where the numerator $2\hbar N(L_i, e_i)$ is the sum of the spin angular momentum magnitudes of the gravitons emitted and $N(L_i, e_i)$ is the total number of gravitons emitted in the gravitational waves during nonrelativistic inspiral from an initial eccentricity $e_i$ down to a final eccentricity $e = 0$ and the denominator $L_i$ is the magnitude of the initial orbital angular momentum.
If the orbit starts off with unit eccentricity $e_i=1$,
we get the value $\nu(1) = 1.002\, 268\, 666\, 2 \pm 10^{-10}$ which confirms the Page's conjecture that the true value of $\nu(1)$ will lie between $1.001\cdots$ and $1.003\cdots$.
We also show that the formula $F(e)$ for gravitons emitted, originally expressed as an infinite series, can be represented
by a single function through an integral representation.
\\


Keywords: Gravitational Wave, Black Hole Binary, Graviton Number

\vspace{1cm}

\today

\end{titlepage}

\renewcommand{\thefootnote}{\arabic{footnote}}
\setcounter{footnote}{0}

\section{Introduction}

In general relativity, gravitational waves are produced by accelerating masses. Gravitational radiation has many aspects analogous to electromagnetic radiation from accelerating charges. A significant difference is that there is no analog to electromagnetic dipole radiation but gravitational radiation at leading order is quadrupolar. The theory of general relativity is a fully non-linear theory, which could make any Newtonian analysis wholly unreliable.
However it turns out \cite{ligo} that the Newtonian approximation describes well a binary system until quite late in its evolution. Then one can calculate
the rate of loss of orbital energy to gravitational waves, when the velocities of the orbiting objects are not close to the speed of light and the strain is not too large \cite{pet-mat,peters}.

The adiabatic approximation, often used in calculations of binary black hole inspirals \cite{pet-mat,peters}, assumes that the changes in the system's parameters (like the semimajor axis and period) are small over a single orbit. While this assumption is valid in the late stages of the inspiral when the orbit is highly circular, it can break down in the early stages when the orbit is highly eccentric. In recent works \cite{page1,page2}, Page has pointed out that discrete orbit effects invalidate the previously widely used adiabatic approximation for the evolution of the nonrelativistic energy, semimajor axis, and period of inspiraling binary black holes. Page argued \cite{page1,page2} that these discrete orbit effects can significantly lengthen the merger time compared to estimates based on the adiabatic approximation. This is especially true for initially highly eccentric orbits and the period of each orbit changes almost entirely only near the periapsis at the beginning of each orbit, where most of the gravitational radiation is emitted.

The paper \cite{page3} provides an exact estimate for the ratio of the sum of magnitudes of the spin angular momentum $2\hbar$ to the magnitude of the total angular momentum vector emitted per orbit during the nonrelativistic inspiral of two black holes or point masses. For two black holes (or other effectively point masses) inspiraling from a highly eccentric nonrelativistic Keplerian orbit with eccentricity $e$, the formula $F(e) = 2 \hbar \frac{dN}{dJ}$ can be found by considering the ratio of the orbit-averaged rate of graviton emission $\frac{dN}{dt}$ to the angular momentum emission rate $\frac{dJ}{dt}$ radiated into gravitational waves averaged over one orbit, which had been computed in \cite{pet-mat,peters}. It was shown in \cite{page3} that the ratio $F(e)$ is remarkably close to the unity. This means that the number of gravitons emitted during binary inspiral is equal (within 1 \%) to the magnitude of the total angular momentum vector emitted divided by the spin angular momentum of a single graviton $(2\hbar)$.

The number of gravitons radiated during a binary inspiral is a complex question that involves quantum gravity 
as well as general relativity.
While we do not have a complete understanding of quantum gravity,
the formula $F(e) = 2 \hbar \frac{dN}{dJ}$ derived from general relativity
may help us understand how binary black holes merge via the emission of gravitational waves.
In this paper, we derive the function $F(e)$ exactly although there
does not seem to be an explicit closed-form elementary function of the eccentricity $e$.
We show that the function $F(e) = 2 \hbar \frac{dN}{dJ}$ is a slowly growing monotonic function of the eccentricity $0 \le e \le 1$ and $F(1) = 1.0128 \cdots $.
We also get an exact result for the ratio
$\nu (e_i)$ of the sum of the spin angular momentum magnitudes of gravitons emitted to
the initial angular momentum of inspiraling binary black holes.
If the orbit starts off with unit eccentricity $e_i=1$,
we get the exact value $\nu(1) = 1.002\, 268\, 666\, 2 \pm 10^{-10}$ which confirms the Page's conjecture \cite{page3}.
It is also shown that the formula $F(e)$ for graviton number
emitted can be determined by a single function using the integral representation of $F(e)$. This integral representation illuminates
why finding a closed form for $F(e)$ is difficult.

\section{Exact results for graviton number emitted}

Consider two black holes (or effectively point masses) with masses $M_1$
and $M_2$, a total mass $M=M_1 +M_2$, reduced mass $\mu = \frac{M_1 M_2}{M_1 + M_2}$, and dimensionless mass ratio $\eta= \frac{\mu}{M}$.
We assume they are slowly inspiraling along a squence of nearly Keplerian orbits, each with semimajor axis $a$, essentricity $e$, period $\tau = 2\pi \sqrt{\frac{a^3}{GM}}$, nonrelativistic energy $E = - \frac{G M \mu}{2a}$,
and angular momentum $L=\sqrt{GM \mu^2 a (1-e^2)}$. We may think of this black hole binary as a hydrogen atom made of two black holes, except that the transition between two energy levels occurs via gravitational radiations instead of electromagnetic radiations. In this paper, $J_n(x)$ denotes
a Bessel function of the first kind of integral order $n$ \cite{watson-book}.

The average gravitational power per orbit radiated into the $n$th harmonic
(angular frequence $\omega_n = n \omega = n \sqrt{\frac{GM}{a^3}}$ ) is \cite{pet-mat,peters}
\begin{equation}\label{power-p}
  P(n) = \frac{32 G^4 M^5 \eta^2}{5 c^5 a^5} g(n,e),
\end{equation}
where
\begin{eqnarray} \label{gne}
  g(n,e) &=& \frac{n^4}{32} \left[ J_{n-2}(ne) - 2e J_{n-1} (ne)
  + \frac{2}{n} J_n (ne) + 2e J_{n+1} (ne) - J_{n+2} (ne) \right]^2 \xx
  && + \frac{n^4}{32} (1-e^2) \left[ J_{n-2}(ne) - 2 J_{n} (ne)
  + J_{n+2} (ne) \right]^2 + \frac{n^2}{24} [ J_{n}(ne)]^2.
\end{eqnarray}
There exists a closed formula for the orbit-averaged gravitational
wave power in all harmonics and it is given by \cite{pet-mat}
\be \label{tot-p}
P = \frac{32 G^4 M^5 \eta^2}{5 c^5 a^5} \sum_{n=1}^\infty g(n,e)
= \frac{32 G^4 M^5 \eta^2}{5 c^5 a^5} \frac{1 + \frac{73}{24} e^2 + \frac{37}{96} e^4  }{(1-e^2)^{\frac{7}{2}}}.
\ee

Since each graviton has energy $\hbar \omega_n = n \hbar \sqrt{\frac{GM}{a^3}}$, the orbit-averaged number rate of graviton emission is
\begin{equation}\label{graviton-n}
  \frac{dN}{dt} = \sum_{n=1}^\infty \frac{P(n)}{\hbar \omega_n}
  = \frac{32 G^{\frac{7}{2}} M^{\frac{9}{2}} \eta^2}{5 \hbar c^5 a^{\frac{7}{2}}} \sum_{n=1}^\infty \frac{g(n,e)}{n}.
\end{equation}
The angular momentum $J$ per time radiated into gravitational waves averaged over one orbit is also known as \cite{peters}
\begin{equation}\label{ang-n}
  \frac{dJ}{dt}  = \frac{32 G^{\frac{7}{2}} M^{\frac{9}{2}} \eta^2}{5 c^5 a^{\frac{7}{2}}} \frac{1 + \frac{7}{8} e^2}{(1-e^2)^2}.
\end{equation}
Therefore, the ratio of the sum of magnitudes of the spin angular momentum $2\hbar$ to the magnitude of the total orbital angular momentum vector emitted per orbit is
\begin{equation} \label{fe}
F(e) \equiv 2 \hbar \frac{dN}{dJ} = \frac{2(1-e^2)^2}{1 + \frac{7}{8} e^2}
\sum_{n=1}^\infty \frac{g(n,e)}{n}.
\end{equation}

Unlike the total power \eq{tot-p} derived from the summation of $g(n,e)$,
a closed form for the function $F(e)$ is not known so far.
But there are many known properties for the function $F(e)$ \cite{page3}:
\bea \la{aymp-fe1}
&& F(e = 0) = 1, \\
\la{aymp-fe2}
&& F(e) = 1 + \frac{1}{48} e^2 - \frac{3}{128} e^4 + {\cal O}(e^6),
\qquad \mathrm{for \; small}\; e < 1, \\
\la{aymp-fe3}
&& F(e \approx 1) = \frac{248}{45 \sqrt{3} \pi} \approx 1.012 \, 811\, 600 \, 479, \qquad \mathrm{for}\; e \approx 1.
\eea
These results are obtained by using the asymptotic expansions of the Bessel functions. We will show that $F(e)$ is a monotonic function
for $0 \le e \le 1$. This means that the function $F(e)$ is remarkably close to the unity for $0 \le e \le 1$. This implies that the number of gravitons emitted during binary inspiral is equal (within 1 \% or perhaps about 0.2 \%) to the magnitude of the total angular momentum emitted divided by the spin angular momentum of a single graviton $(2\hbar)$.

The conservation of the total angular momentum of the binary black holes and gravitons emitted implies that the decrease of the angular momentum of the orbiting black holes is equal to the sum of the spin and orbital angular momentum vectors of gravitons emitted  averaged over a few wavelengths:
\be \la{angular-cons}
- \Delta \overrightarrow{L}_{BH} = \overrightarrow{L} + \overrightarrow{S} = \overrightarrow{J}, 
\ee
where $\Delta \overrightarrow{L}_{BH}$ is the change of the orbital angular momentum of the binary black hole due to the emissions of graviational waves and $\overrightarrow{L} \; (\overrightarrow{S})$ is the expectation value of the orbital angular momentum operator (the intrinsic spin contribution) carried away by the gravitational wave. See Eqs. (2.51), (3.97) and (3.99) in Ref. \cite{gw-magg} for explicit expressions of $\overrightarrow{L}$ and $\overrightarrow{S}$ and their time derivatives.
Some of the spins of gravitons emitted can be parallel to the orbital
angular momentum, and some can be opposite.
Thus the sum of magnitudes of the spin angular momenta would generically be greater than the magnitude of the sum of the spin angular momentum vectors averaged over one orbit, i.e.,
\be \la{spin-ineq}
\mathcal{S} = |\sum_{i=1}^N \langle \overrightarrow{S}_i \rangle| \leq
\sum_{i=1}^N \langle |\overrightarrow{S}_i| \rangle = 2 \hbar N.
\ee
It is known \cite{pet-mat,peters} that, in the quadrupole approximation, the dominant type of radiation is ``magnetic quadrupole" with $J=2$. The magnetic quadrupole is defined using the tensor spherical harmonics $T_{JLM} (\Omega)$ which represents the combination of an orbital angular momentum $\overrightarrow{L}$ and a spin angular momentum $\overrightarrow{S} \; (S=2)$ to form a total angular momentum $\overrightarrow{J}$ (see Eqs. (7)-(9) in \cite{pet-mat}).\footnote{The dimensionless quantum numbers $(J,L,M)$ refer to eigenvalues of operators $(\overrightarrow{J}^2, \overrightarrow{L}^2, J_z)$ acting on the spherical harmonics $T_{JLM} (\Omega)$ such that $\overrightarrow{J}^2 T_{JLM} (\Omega) = J(J+1) \hbar^2 T_{JLM} (\Omega), \;  \overrightarrow{L}^2 T_{JLM} (\Omega) = L(L+1) \hbar^2 T_{JLM} (\Omega)$, and $J_z T_{JLM} (\Omega) = M \hbar T_{JLM} (\Omega)$.} In this limit, the $z$-component of angular momentum $J_z = \overrightarrow{J} \cdot \hat{\mathbf{k}}
= \overrightarrow{S} \cdot \hat{\mathbf{k}} = S_z$ for a propagation direction $\hat{\mathbf{k}}$ can take values $M = -2, 0, 2$. Emission of an $M=2$ graviton would decrease the orbital angular momentum of the black holes by $2 \hbar$, an $M=0$ graviton would not change it at all and an $M = - 2$ graviton would increase the orbital angular momentum of the black holes by $2 \hbar$. Therefore, in the quadrupole approximation where there is a non-zero amplitude for $M =-2$ (when $e > 0$), the decrease \eq{angular-cons} in the black hole orbital angular momentum will be less than the number of gravitons multiplied by $2 \hbar$ (the magnitude of the spin of each graviton).
It may explain why $F(e) > 1$ when $e >0$.\footnote{We thank Don Page for bringing this discussion to our attention.}

\begin{figure}
\begin{center}
\includegraphics[width=0.6\linewidth]{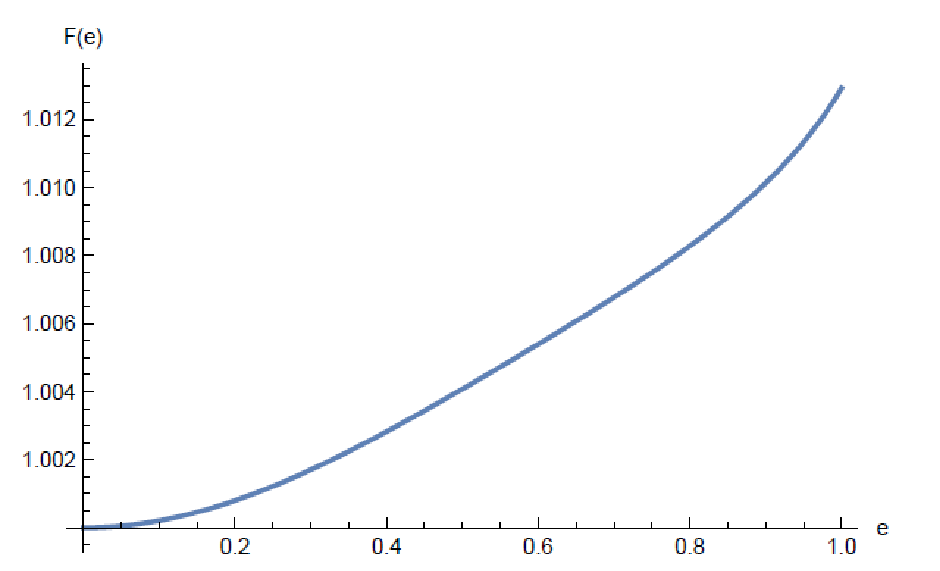}
\end{center}
\caption{The graph of the function $F(e)$ with $N=100$. $N \; (1 \le n \le N)$ denotes a truncation of the summation into finite terms, i.e. $\sum_{n=1}^\infty \to \sum_{n=1}^N$.}
\label{fig_F(e)}
\end{figure}

\begin{figure}
    \centering
    \includegraphics[width=0.5\linewidth]{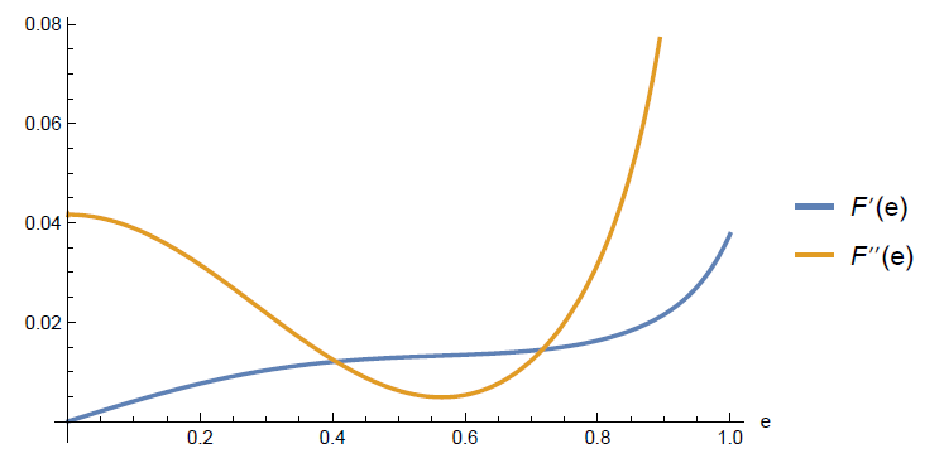}
    \caption{The graphs of the functions $F'(e) = \frac{dF(e)}{de}$
    and $F^{''}(e) = \frac{d^2 F(e)}{de^2}$ with $N=100$.}
    \label{f1-f2}
\end{figure}

Therefore we will compute the function $F(e)$ exactly although there
does not seem to be an explicit closed-form elementary function of the eccentricity $e$. Later we will illuminate why finding a closed form for $F(e)$ is difficult. We will use Mathematica and the add-on
package MathSymbolica for the calculation of the summation in \eq{fe}.
We find the result
\bea \la{ge-exact}
G(e) &\equiv& \sum_{n=1}^\infty \frac{g(n,e)}{n} = \frac{1}{2 (1 - e^2)^2} \times \\
&& \sum_{n=1}^\infty e^{2n-4} \left( \left(1 - 5 e^2 + \frac{43}{3} e^4
- n(1-e^2)(4-11e^2) + 2n^2 (1-e^2)^2 \right) c_{2n}
+ (1-e^2)^2 d_{2n} \right), \nonumber
\eea
where
\bea \la{cn}
c_{2n} &=& \frac{ \left(n- \frac{5}{2} \right) !}{\sqrt{\pi} (n-2)!}
\sum_{s=0}^{n-3} \frac{(-1)^s (n-s-2)^{2n-3}}{(2n-s-4)! s!}
- \frac{2 \left(n- \frac{3}{2} \right) !}{\sqrt{\pi}(n-1)!}
\sum_{s=0}^{n-2} \frac{(-1)^s (n-s-1)^{2n-1}}{(2n-s-2)! s!} \xx
&& + \frac{ \left(n - \frac{1}{2} \right) !}{\sqrt{\pi} n!}
\sum_{s=0}^{n-1} \frac{(-1)^s (n-s)^{2n+1}}{(2n-s)! s!}, \\
\la{dn}
d_{2n} &=& \frac{(n-2) \left(n-\frac{5}{2} \right) !}{\sqrt{\pi} (n-1)!}
\sum_{s=0}^{n-3} \frac{(-1)^s (n-s-2)^{2n-3}}{(2n-s-4)! s!}
- \frac{2 \left(n - \frac{1}{2} \right) !}{\sqrt{\pi} n!}
\sum_{s=0}^{n-2} \frac{(-1)^s (n-s-1)^{2n-1}}{(2n-s-2)! s!} \xx
&& + \frac{\left(n - \frac{1}{2} \right) !}{\sqrt{\pi} n!}
\sum_{s=0}^{n-1} \frac{(-1)^s (n-s)^{2n+1}}{(2n-s)! s!}.
\eea
The coefficients satisfy the property
\be \la{sum-rule}
\sum_{n=1}^\infty c_{2n} = \frac{1}{\sqrt{3} \pi}, \qquad
\sum_{n=1}^\infty d_{2n} = \frac{2}{\sqrt{3} \pi}.
\ee
The expansion \eq{ge-exact} is exact and so valid for $0 \le e \le 1$.
Thus one can use a computational software system like Mathematica to calculate \eq{ge-exact} up to a desired order.
The result for lower powers of $e$ is given by
\bea \la{exp_gefe}
G(e) &=& \frac{1}{2} + \frac{139}{96} e^2 + \frac{919}{384} e^4
+ \frac{61583}{18432} e^6
+ \frac{1897649}{442368} e^8 + \frac{926986067}{176947200} e^{10}
+ \frac{2627868443}{424673280} e^{12} + \cdots, \xx
F(e) &=& 1 + \frac{1}{48} e^2 - \frac{3}{128} e^4
+ \frac{59}{2304} e^6 - \frac{4603}{221184} e^8 + \frac{558239}{29491200} e^{10} - \frac{34385443}{2123366400} e^{12} + \cdots.
\eea
The result in \eq{exp_gefe} precisely reproduces Eq. (37) and Eq. (38)
in \cite{page3} up to $e^4$-order.
The above result already shows that $G(e)$ is divergent when $e \to 1$
while $F(e)$ is convergent. It turns out \cite{page3} that the asymptotic behavior
of $G(e)$ is $G(e) \sim \frac{31}{6 \sqrt{3} \pi} (1-e^2)^{-2}$
when $e \to 1$. We found that $F(1) = 1.012\, 815 \, 967 \,259 \,511$ up to $N=100$, which is very close to the value in \eq{aymp-fe3}.
The graph of the function $F(e)$ is shown in Fig. \ref{fig_F(e)}.
The figure shows that the function $F(e)$ is a slowly increasing monotonic function of $e$ in $[0,1]$. In order to test the monotonicity of the function $F(e)$ more concretely, we examined the behaviors of the functions $F'(e) = \frac{dF(e)}{de}$ and $F^{''}(e) = \frac{d^2 F(e)}{de^2}$ up to $e^{40}$-order. Fig. \ref{f1-f2} clearly shows that the function $F(e)$ is monotonic.
In Fig. \ref{f1-f2}, $F''(e)$ appears to increase sharply near $e=1$ (due to a scale), but it turns out that
\be \la{ddF(e=1)}
F'' (e=1) = \frac{17056}{10125 \sqrt{3} \pi} + \frac{32}{135} \sum_{n=1}^\infty (2n + 3n^2) c_{2n} \approx 0.316259.
\ee
Thus the function $F'' (e)$ is finite, not divergent, near $e=1$.

The ratio of the orbital angular momentum $L(e)$ when the eccentricity is $e$, to $L_1$ at eccentricity $e=1$ if the inspiral had started there, can be found using Eq. (5.11) of Ref. \cite{peters} to be
\be \la{lambda-e}
\lambda(e) \equiv \frac{L(e)}{L_1} = e^{\frac{6}{19}}
\left( \frac{304 + 121 e^2}{425} \right)^{\frac{435}{2299}}.
\ee
If an inspiral starts at initial orbital angular momentum $L_i$ and eccentricity $e_i$, the orbital angular momentum during the inspiral as the eccentricity decreases below $e_i$ is
\be \la{l_e}
L(e) = L_i \frac{\lambda (e)}{\lambda (e_i)}.
\ee
In order to calculate the total number of gravitons
\be \la{totn-graviton}
N (L_i, e_i) = \frac{L_i}{2 \hbar} \nu(e_i)
\ee
emitted during the inspiral from initial orbital angular momentum $L_i$ and eccentricity $e_i$ to final eccentricity very close to zero,
one can use the results \eq{fe} and \eq{lambda-e}. One can calculate the number of gravitons as \cite{page3}
\be \la{N-graviton}
N (L_i, e_i) = \frac{L_i}{2 \hbar} \nu(e_i) = \frac{L_i}{2 \hbar \lambda(e_i)}
\int_0^{e_i} F(e) \frac{d\lambda}{de} de.
\ee

\begin{figure}
    \centering
    \includegraphics[width=0.5\linewidth]{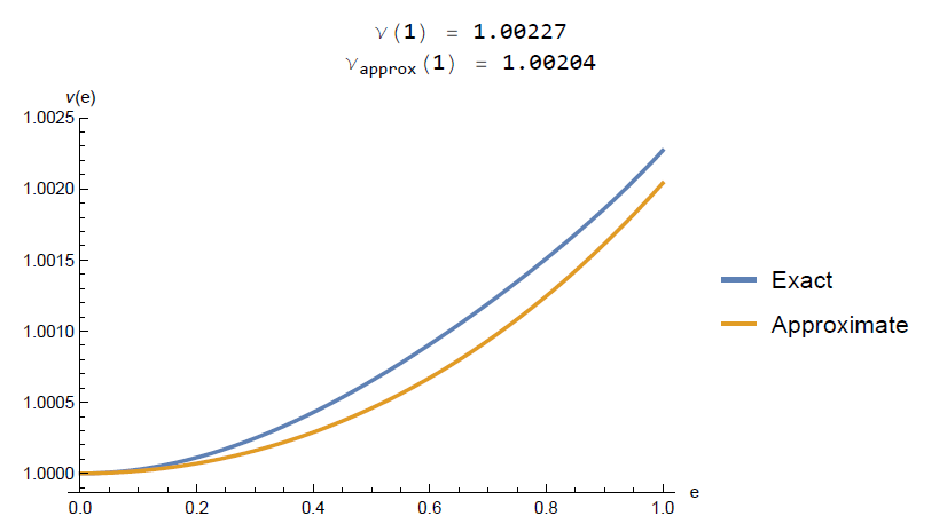}
    \caption{$\nu(e_i)$ with $N=100$. The blue graph is the result calculated using \eq{exact-nu},
    and the orange graph is the result calculated using Eq. (33) in \cite{page3}.}
    \label{n(1)-ours and page}
\end{figure}

We can derive the exact result given by
\bea \la{exact-nu}
&& \nu (e_i) =  \frac{2 \hbar N(L_i, e_i)}{L_i} =
\frac{1}{\lambda(e_i)} \int_0^{e_i} F(e) \frac{d\lambda}{de} de =  \xx
&& \;\; \left( \frac{304}{425} \right)^{\frac{435}{2299}}
\frac{1}{\lambda(e_i)} \left[ \frac{272}{553 e_i^{\frac{32}{19}}}
\left( \left(1 + \frac{1501}{408} e_i^2 \right) \left(1 + \frac{121}{304} e_i^2 \right)^{\frac{435}{2299}} -
\setlength\arraycolsep{1pt}
{}_2 F_1 \left(- \frac{16}{19}, \frac{1864}{2299}; \frac{3}{19};
-\frac{121}{304} e_i^2 \right) \right) \right.  \xx
&& \hspace{1.5cm} + \sum_{n=2}^\infty e_i^{ 2n - \frac{70}{19}} \left\{ c_{2n}
\left( 3 \frac{2n^2 -4n +1}{19n-35}
\setlength\arraycolsep{1pt}
{}_2 F_1 \left( \frac{1864}{2299}, n - \frac{35}{19}; n- \frac{16}{19};
-\frac{121}{304} e_i^2 \right) \right. \right. \xx
&& \hspace{1.5cm} - 3 \frac{4n^2 - 15n + 5}{19n-16} \;
\setlength\arraycolsep{1pt}
{}_2 F_1 \left( \frac{1864}{2299}, n - \frac{16}{19}; n + \frac{3}{19};
-\frac{121}{304} e_i^2 \right)  e_i^2 \xx
&& \hspace{1.5cm} \left. + \frac{6n^2 - 33n + 43}{19n+3} \;
\setlength\arraycolsep{1pt}
{}_2 F_1 \left( \frac{1864}{2299}, n + \frac{3}{19}; n + \frac{22}{19};
-\frac{121}{304} e_i^2 \right) e_i^4 \right) \xx
&& \hspace{1.5cm} + 3 d_{2n} \left( \setlength\arraycolsep{1pt}
\frac{1}{19 n- 35} \; {}_2 F_1 \left( \frac{1864}{2299}, n - \frac{35}{19}; n - \frac{16}{19};
-\frac{121}{304} e_i^2 \right) \right. \xx
&& \hspace{1.5cm} - \frac{2}{19 n- 16} \; {}_2 F_1 \left( \frac{1864}{2299}, n - \frac{16}{19}; n + \frac{3}{19};
-\frac{121}{304} e_i^2 \right) e_i^2   \xx
&& \hspace{1.5cm} \left. \left. \left. + \frac{1}{19 n + 3} \; {}_2 F_1 \left( \frac{1864}{2299}, n + \frac{3}{19}; n + \frac{22}{19};
-\frac{121}{304} e_i^2 \right)  e_i^4 \right) \right\} \right],
\eea
where ${}_2 F_1 (a,b;c;x)$ is a hypergeometric function \cite{arfken}
and the coefficients $c_{2n}$ and $d_{2n}$ are given by \eq{cn} and \eq{dn}, respectively. Although the formula takes a rather complex form, it is straightforward to calculate \eq{exact-nu} up to a desired order using a computational software system like Mathematica.
Page conjectured the true value $\nu(1)$ to lie between $1.001\, 219\, 302\, 812$ and $1.003\, 327\, 748\, 211$ \cite{page3}. For the calculation, Page approximated Eq. \eq{lambda-e} with finite terms up to $g^5$ where $g = \lambda^{\frac{19}{3}}$. We can get the exact value for $\nu(1) = 1.002\, 268\, 666\, 2 \pm 10^{-10}$ when $e_i \to 1$, using the result \eq{exact-nu}. So we confirm the Page's conjecture.
For comparison, our exact result and the result of Eq. (33) in \cite{page3}
are shown in Fig. \ref{n(1)-ours and page}.

\section{Integral representation of $F(e)$}

Using the recurrence relations and Bessel's equation, $g(n,e)$ in \eq{gne} can be rewritten as \cite{pet-mat,page3}
\bea \la{re-gne}
2 e^4 \frac{g(n,e)}{n} &=&
(1-e^2)^3  n^3 J_{n}^2 (ne) + \left( 1 - e^2 + \frac{1}{3} e^4 \right)
n J_n^2 (ne) - e (1-e^2) (4-3e^2) n^2 J_{n} (ne) J'_{n} (ne) \xx
&& + e^2 (1-e^2) \left( (1-e^2) n^3 + n \right) J^{'2}_{n}(ne),
\eea
where the prime means the derivative with respect to the argument $x \equiv ne$,
i.e. $' \equiv \frac{d}{dx}$.
Therefore, the function $G(e)$ in \eq{ge-exact} can be written as
\bea \la{re-ge}
G(e) &=& \frac{1}{2 e^4} \left( (1-e^2)^3  S_1 (e)
+ \left( 1 - e^2 + \frac{1}{3} e^4 \right) P_1 (e) \right. \xx
&& \left. - \frac{1}{2}e (1-e^2) (4-3e^2) \frac{dP_1 (e)}{de}
+ e^2 (1-e^2)^2 S_2 (e) + e^2 (1-e^2) P_2 (e) \right),
\eea
where
\bea \la{sp12}
&& S_1 (e) = \sum_{n=1}^\infty n^3 J_{n}^2 (ne), \qquad
S_2 (e) = \sum_{n=1}^\infty n^3 J^{'2}_{n}(ne), \xx
&& P_1 (e) = \sum_{n=1}^\infty n J_n^2 (ne), \qquad
P_2 (e) = \sum_{n=1}^\infty n J^{'2}_{n}(ne).
\eea

Differentiate $S_2 (e)$ with respect to $e$ to obtain
\be \la{der-s2}
\frac{d S_2(e)}{de} = 2 \sum_{n=1}^\infty n^4 J'_{n}(ne) J^{''}_n (ne).
\ee
Use the Bessel's equation for $J_n (ne)$ in the form
\be \la{bessel-eq}
J^{''}_n (ne) = \frac{1}{e^2} \left( - \frac{e}{n} J'_n (ne)
+ (1-e^2) J_n (ne) \right)
\ee
and substitute \eq{bessel-eq} in \eq{der-s2} to get
the differential equation \cite{lerc-taut}
\be \la{diff-s12}
\frac{d}{de} \left( e^2 S_2 (e) - (1-e^2) S_1 (e) \right) = 2e S_1 (e).
\ee
Then we have
\be \la{s2-s1}
S_2 (e) = \frac{1}{e^2} \left( (1-e^2) S_1(e) + 2 \int^e_0 dx x S_1 (x) \right),
\ee
where we used the fact that the integration constant in \eq{s2-s1} is zero
by evaluation as $e \to 0$.

To evaluate $S_1(e)$, note that (see {\bf 5.43} \, (1) in \cite{watson-book}
and {\bf 6.681} \, 1. in \cite{grd-ryz-book} )
\be \la{jj-int}
J_n^2 (ne) = \frac{2}{\pi} \int_0^{\frac{\pi}{2}} J_{2n} (2ne \cos \theta) d \theta,
\ee
which gives us the integral representation of $S_1 (e)$
\be \la{int-s1}
S_1 (e) = \frac{1}{4\pi} \int_0^{\frac{\pi}{2}} \sum_{n=1}^\infty (2n)^3 J_{2n} (2ne \cos \theta) d \theta.
\ee
The summation in \eq{int-s1} is known as a Kapteyn series. See Chap. XVII in \cite{watson-book}. If the Kapteyn series
\be \la{kap-f}
f(z) = \sum_{m=1}^\infty a_m J_m (mz)
\ee
is known, then the series
\be \la{kap-F}
F(z) = \sum_{m=1}^\infty \frac{a_m}{m^2} J_m (mz)
\ee
is given by two simple integrations because
\be \la{kap-rel}
{\cal L}_z F(z) \equiv \frac{1}{1-z^2} \left( z \frac{d}{dz} \right)^2 F(z) = f(z)
\ee
by direct differentiation of \eq{kap-f}. See {\bf 17.33} in \cite{watson-book}
and \cite{nikishov} for more details.

Assume that, in Eq. \eq{kap-f},
\be \la{am}
a_m = \left\{
  \begin{array}{ll}
    m^3, & \hbox{$m =$ even;} \\
    0, & \hbox{$m=$ odd.}
  \end{array}
\right.
\ee
Then we have
\be \la{f-F}
f(z) = \sum_{n=1}^\infty (2n)^3 J_{2n} (2nz) = \left({\cal L}_z \circ {\cal L}_z \right) Q(z)
\ee
where
\be \la{qe}
Q(z) = \sum_{n=1}^\infty \frac{1}{2n} J_{2n} (2nz).
\ee
Therefore, the function $S_1 (e)$ in \eq{int-s1} is given by
\be \la{intrep-s1}
S_1 (e) = \frac{1}{4\pi} \int_0^{\frac{\pi}{2}} \left({\cal L}_z \circ {\cal L}_z \right) Q(z) d\theta
\ee
where $z = e \cos \theta$. Once the integral \eq{intrep-s1} is calculated,
$S_2 (e)$ can also be determined through \eq{s2-s1}. Since $S_1(e)$ will be given as a series expansion in $e$, computing the integral in \eq{s2-s1} is straightforward. We can apply a similar strategy to calculate $P_1(e)$ and $P_2 (e)$ in \eq{sp12}. The result is
\bea \la{int-p1p2}
&& P_2 (e) = \frac{1}{e^2} \left( (1-e^2) P_1(e) + 2 \int^e_0 dx x P_1 (x) \right), \xx
&& P_1 (e) = \frac{1}{\pi} \int_0^{\frac{\pi}{2}} {\cal L}_z Q(z) d\theta.
\eea

In consequence, the function $G(e)$ in \eq{re-ge} and hence the function $F(e)$ in \eq{fe} can be completely determined as long as
the function $Q(z)$ in \eq{qe} is known. Using the integral representation of the Bessel function (see {\bf 8.411} \, 1. in \cite{grd-ryz-book})
\be \la{intrep-bess}
J_n(n z) = \frac{1}{2\pi} \int_{-\pi}^\pi e^{-in (\theta - z \sin \theta)} d\theta,
\ee
the function $Q(z)$ in \eq{qe} can be represented by the integral
\be \la{intrep-qz}
Q(z) = - \frac{1}{2\pi} \int_{0}^\pi \ln \left(2 \sin (\theta
- z \sin \theta) \right) d\theta
\ee
which is equal to (3.45) in \cite{nikishov}. The integral of $Q(z)$ is not known in a closed form, so the best way is to use the power series expansion for the Bessel function in Eq. \eq{qe}:
\bea \la{qz-int}
Q(z) &=& \frac{1}{2} \sum_{n=1}^\infty z^{2n}
\sum_{s=0}^{n-1} \frac{(-1)^s (n-s)^{2n-1}}{(2n-s)! s!} \xx
&=& \frac{z^2}{4} + \frac{z^4}{12} + \frac{11 z^6}{240} + \frac{151 z^8}{5040} + \frac{15619 z^{10}}{725760} + \cdots.
\eea
The second line in \eq{qz-int} is exactly the same as (3.50) in \cite{nikishov}.
It is straightforward to reproduce the result \eq{qz-int} by
expanding the integrand in \eq{intrep-qz} as a power series in $z$
and then performing the integration.
The integral representation \eq{intrep-qz} clarifies the reason
why finding a closed-form elementary function for $F(e)$ is difficult.
It is straightforward to calculate $G(e)$ in \eq{re-ge} and reproduce $F(e)$ in \eq{fe} using the result \eq{qz-int}.

\section{Discussion}

The momentum conservation forbids dipole radiation in the case of gravitational radiation. Hence, at leading order, gravitational radiation is quadrupolar. Suppose that a binary system consists of non-spinning objects.
The luminosity radiated as gravitational waves during the evolution of the binary system is given by \eq{tot-p}, which was averaged
over one orbit. The energy loss through the gravitational wave emission
drains the orbital energy of the binary $E = - \frac{G\mu^2}{2a}$, thus the energy conservation implies that $\frac{dE}{dt} = - P$.
Therefore the orbit of a binary system decays toward a final coalescence
under the dissipative effect of gravitational radiation reaction.

Gravitational waves, as any field, carry away a total angular momentum
$\overrightarrow{J}$ that is made of a spin contribution and of an orbital angular momentum contribution.
According to Eq. \eq{angular-cons}, this total angular momentum is drained from the total angular momentum 
of the source which, for a macroscopic source, is a purely orbital angular momentum \cite{gw-magg}.
In section 2, we have defined the magnitude of the sum of the spin angular momenta of gravitons as $\mathcal{S}\leq 2 \hbar N$ where $N$ is the number of gravitons emitted during inspiraling orbital motion.
For a circular orbit with $e=0$, one can see using \eq{graviton-n} and $G(e=0) = \frac{1}{2}$ that the equality 
$\mathcal{S} = 2 \hbar N$ holds.
However Page observed \cite{page3} that it is not true for a generic eccentricity $0 < e < 1$.
In this paper we have shown that the function
$F(e) = 2 \hbar \frac{dN}{dJ}$ is a slowly growing monotonic function of the eccentricity $e$
and $F(1) = 1.012\, 815 \, 967 \,259 \,511 \cdots $. Also we have exactly calculated $\nu(e_i) = \frac{2 \hbar N (L_i, e_i)}{L_i}$
for $0 \le e_i \le 1$ and we proved the Page's conjecture
by showing precisely $\nu(1) = 1.002\, 268\, 666\, 2 \pm 10^{-10}$.
We have briefly explained why there is a very small deviation ($\sim$ 0.227 \%) from the unity for the value $F(e)$ when $e > 0$.

In a binary system bound by gravity, the Newtonian approximation of its dynamics will break down as the relative orbital velocity approaches
the speed of light. Corrections to the Newtonian dynamics can be expanded in powers of the post-Newtonian (PN) parameter $x =\frac{v^2}{c^2} = \frac{GM}{c^2 r_{\mathrm{sep}}}$,
where $r_{\mathrm{sep}}$ is an instantaneous orbital separation \cite{pn1-wawi,pn12-app}.
The quadrupole formula can be extended to the first post-Newtonian (1PN) order by analyzing the gravitational waveform
and energy flux in a weak-field limit, far from a binary system.
For a binary inspiral, post-Newtonian corrections to the quadrupole formula scale as powers of $x =\frac{v^2}{c^2}$
with $x$ typically around $10^{-2}$ \cite{pn12-app}.
Then the luminosity $P$ in \eq{tot-p} and the angular momentum change $\frac{dJ}{dt}$ in \eq{ang-n} will get corrections of the order of $10^{-2}$
in 1PN order. The property $F(e)\geq 1$ is expected to continue
even for higher-order PN approximations since the conservation of the total angular momentum \eq{angular-cons} still holds true for  high-order corrections and one can apply a similar argument in Eqs. \eq{angular-cons} and \eq{spin-ineq} to them.
If the property $F(e) \geq 1$ persists in higher-order PN approximations, the conservation of total angular momentum \eq{angular-cons} implies that corrections to the angular momentum $J$ at the 1PN level would be more significant than those to the luminosity $P$. This is because higher-order multipole radiations, though suppressed by a factor of $10^{-2}$ compared to the quadrupole radiation, correspond to higher spin modes.
So it will be interesting to see whether the correction by higher-order terms further strengthens the tendency for $F(e)\ge 1$.

The exact formulae we have derived in this paper represent a leading-order approximation.
Investigating higher-order corrections, including those arising from post-Newtonian effects and spin-orbit/spin-spin interactions \cite{pn1-wawi,pn12-app}, would improve the accuracy of the predictions.
Calculating higher-order PN corrections would be particularly valuable.
Such calculations could provide some evidence for the quantization of gravitational waves in units of $2\hbar$ (graviton spin).

\newpage

\section{Note Added}

After this paper was posted on the arXiv, Don Page reviewed our paper carefully and sent us very helpful comments, pointing out an embarracing error in Eq. \eq{re-gne}. The prime in Eq. \eq{re-gne} should mean the derivative with respect to the argument $x=ne$, not $e$.
He also recommended clarifying the behavior of the function $F''(e)$ near $e = 1$ in Fig. \ref{f1-f2}.
So we have included Eq. \eq{ddF(e=1)}. Furthermore, he conjectured a more convenient integral representation for the function $Q(z)$
in Eq. \eq{intrep-qz} based on some data available at https://oeis.org/. His proposed formula was
\be \la{page-qz}
Q(z) = - \frac{1}{2\pi} \int_0^\infty \ln \left( 1 - z^2 \frac{\sin^2 x}{x^2} \right) dx.
\ee
We proved his conjecture by directly computing the integral in this formula:
\bea \la{proof-page}
Q(z) &=& \frac{1}{2\pi} \sum_{n=1}^\infty \frac{z^{2n}}{n}
\int_0^\infty \left( \frac{\sin^{2} x}{x^{2}} \right)^n dx \xx
&=& \frac{1}{2} \sum_{n=1}^\infty z^{2n}
\sum_{s=0}^{n-1} \frac{(-1)^s (n-s)^{2n-1}}{(2n-s)! s!},
\eea
where we used the result
\begin{equation*}
\int_0^\infty \left( \frac{\sin^{2} x}{x^{2}} \right)^n dx = n \pi
\sum_{s=0}^{n-1} \frac{(-1)^s (n-s)^{2n-1}}{(2n-s)! s!}.
\end{equation*}
The result \eq{proof-page} is precisely equal to Eq. \eq{qz-int}.

\section*{Acknowledgments}

We would like to express our sincere gratitude to Don Page for his insightful discussions and clarification of certain points. Additionally, we appreciate his valuable input regarding the angular momentum change in the quadrupole approximation,
the behavior of the function $F''(e)$ near $e = 1$ and suggesting the integral representation
for the function $Q(z)$ in \eq{page-qz}. We also thank Chunglee Kim for helpful discussions.
This research was performed using Mathematica (www.wolfram.com) and the add-on
package MathSymbolica (www.mathsymbolica.com).
This work was supported by the National Research Foundation of Korea (NRF) with grant number NRF-2018R1D1A1B0705011314.



\end{document}